\documentclass[12pt]{article}
\usepackage{amsmath}
\usepackage{graphicx}
\begin{document}
\vspace*{1cm}
\begin{center}
 {\Huge\bf
The theory of Multiverse, multiplicity of physical objects and
physical constants
\vspace*{0.25in}

\large

Alexander K.\ Gouts
\vspace*{0.15in}

\normalsize

Department of Computer Science\\ Omsk State University \\
644077 Omsk-77 RUSSIA
\\
\vspace*{0.3cm}
E-mail: guts@univer.omsk.su  \\
\vspace*{0.5cm}
October 17, 2002\\
\vspace{.3in}
}
ABSTRACT
\begin{quote}
The Multiverse is collection of parallel universes.
In this article a formal theory and a topos-theoretic models of the
multiverse are given. For this the Lawvere-Kock Synthetic
Differential Geometry
and topos  models for smooth infinitesimal analysis are used.
Physical properties of multi-variant and many-dimensional parallel
universes are discussed. The source of multiplicity of physical objects is set of
physical constants.
\end{quote}
\end{center}

\vfill

This paper contains the report on 11-th International (Russian) Gravitational Conference
at Tomsk. July 1-7, 2002.
\newpage
\def\R{{{\rm I} \! {\rm R}}}
\def\C{{{\rm I} \! \hspace*{-1.3mm} {\rm C}}}
\def\L{{{\rm I} \! {\rm L} }}
\def\D{{{\cal D} }}
\def\M{{{\cal M} }}
\def\F{{{\cal F} }}
\def\G{{{\cal G} }}
\def\Z{{{\cal Z} }}
\def\T{{{\cal T} }}
\def\d{{{\rm \Delta} \!\!\!\! {\Delta}}}
\def\N{{{\rm I} \! {\rm N} }}
\def\r{{{\cal R} }}

\section{Intoduction}

In the  Deutsch 's book \cite{1} the sketch of structure of  physical
reality  named Multiverse which is set of the parallel universes is given.
Correct description of the Multiverse
can be done only within the framework of the quantum theory.

In this article a sketch of formal theory and topos-theoretic models of the Deutsch
multiverse are given (see more in \cite{A1}).

We wish to preserve the framework of the mathematical apparatus
of the 4-dimensional General theory of Relativity,
and so we shall consider the Universe as concrete 4-dimensional
Lorentz manifold $<~\r^{4}, g^{(4)}>$ (named space-time).

\section{Formal theory of Multiverse}

We construct the theory of Multiverse as formal theory $\T$ which
is maximally similar to
 the General theory of Relativity, i.e. as theory of {\it one}
4-dimensional universe,
 but other parallel universes must
appear under costruction of models of formal theory.

The basis of our formal theory  $\T$ is the Kock-Lawvere
Synthetic Differential Geometry (SDG) \cite{2}.

It is important to say that SDG has no any set-theoretic model because Lawvere-Kock axiom
is  incompatible with Law of  excluded middle.
Hence we shall construct formal theory of Multiverse
on base of the intuitionistic logic. Models for this theory
are smooth topos-theoretic models and for stheir description the
usual classical logic is used.

In SDG the commutative ring $\r$ is used instead of real field
 $\R$. The ring $\r$ must satisfy the following

\medskip\noindent
{\bf Lawvere-Kock axiom}. {\it Let $D=\{x\in \r : x^2=0\}$. Then
  $$
  \forall(f\in \r^D)  \exists ! (a,b)\in \r\times \r\ \forall
d\in D( f(d)=a+b\cdot d).
  $$
}
\medskip
and some other axioms (see in \cite[Ch.VII]{3}.).

\smallskip

Ring  $\r$ includes real numbers from $\R$ and has new elements named
{\it infinitesimals} belonging to "sets"
$$
D=\{d\in \r : d^2=0\},..., D_k=\{d\in \r : d^{k+1}=0\},...
$$
$$
\d=\{x\in \r : f(x)=0, \ \mbox{all}\ f\in m_{\{0\}}^g \},
$$
where $m^g_{\{0\}}$ is ideal of smooth functions having zero germ  at 0,
i.e. vanishing
in a neighbourhood of 0.

We have
$$
D\subset D_2\subset ... \subset D_k\subset... \subset \d.
$$

We can construct
Riemmanian geometry
for four-dimensional (formal) manifolds $<\r^4, g^{(4)}>$. These
manifolds are basis
for the Einstein theory of gravitation \cite{4}.

\smallskip
We postulate that {\it multiverse is four-dimensional space-time in SDG,
i.e. is a formal Lorentz manifold $<\r^4, g^{(4)}>$ for which
the Einstein field equations are held:
\begin{equation}\label{ein}
R^{(4)}_{ik}-\frac{1}{2}g^{(4)}_{ik}(R^{(4)}-2\Lambda)=
\frac{8\pi G}{c^4}T_{ik}.
\end{equation}
}

\smallskip
A solution of these equations is 4-metric $g^{(4)}(x),$ $x\in \r$.

\smallskip

Below we consider the physical consequences of our theory in so
called well-adapted
smooth topos models of the form ${\bf Set}^{\L^{op}}$ which contain as full subcategory
the category of smooth manifolds $\M$.

\section{Smooth topos models of multiverse}

Let  $\L$ be  dual category for category of finitely generated
$C^\infty$-rings.
It is called {\it category of loci} \cite{3}. The objects of
$\L$ are  òû ¦ª¸ 
finitely generated $C^\infty$-rings, and morphisms are reversed morphisms of
category of finitely generated $C^\infty$-rings.

The object (locus) of $\L$ is denoted as $\ell A$, where  $A$ is
a $C^\infty$-ring.
Hence, $\L$-morphism $\ell A \to \ell B$ is $C^\infty$-homomorphism $B \to A$.

A finitely generated $C^\infty$-ring $\ell A$ is isomorphic to ring of the
form $C^\infty(\R^n)/I$ (for some natural number $n$ and
some ideal $I$ of finitely generated functions).

Category  ${\bf Set}^{\L^{op}}$ is topos \cite{3}. We consider topos
${\bf Set}^{\L^{op}}$ as model of formal theory of multiverse.

\smallskip
With the Deutsch point of view  the transition to concrete model
of formal theory is creation of {\it virtual reality} \footnote{This  thought
belongs to Artem Zvyagintsev.}. Physical Reality that we  perceive was called
by Deutsch {\it Multiverse}
 \footnote{Multiverse = many (multi-) worlds;
 universe is one (uni) world. }. Physical Reality is also
virtual reality which was created our brain
\cite[p.140]{1}.

\medskip
A model of multiverse is {\it generator of virtual reality} which
has some {\it repertoire of  environments}.
Generator of virtual reality creates environments and we
observe them.  Explain it.

\medskip
Under interpretation $i: {\bf Set}^{\L^{op}}\models \T$ of formal
multiverse theory $\T$ in topos ${\bf Set^{\L^{op}} }$
the objects of theory, for example, ring $\r$, power $\r^\r$ and so on are
interpreted
as objects of topos, i.e.
functors $F=i(\r)$, $F^F=i(\r^\r)$ and so on. Maps, for example,
$\r\to \r$,  $\r\to \r^\r$ are now morphisms of topos
 ${\bf Set}^{\L^{op}}$, i.e. natural transformations of functors:
$F\to F$, $F\to F^F$.

Finelly, under interpretation of language of formal multiverse theory
we must interpret elements of ring
$\r$ as "elements" of functors $F\in {\bf Set}^{\L^{op}}$.   In other words
we must give interpretation for relation $r\in \r$. It is very
difficult problem because functor $F$ is defined on category of loci $\L$;
its independent variable is arbitrary locus $\ell A$, and
dependent variable is a set $F(\ell A)\in {\bf Set}$. To solve this problem
we consider {\it generalized elements} $x\in_{\ell A}F$ of functor $F$.

Generalized element $x\in_{\ell A}F$, or {\it element  $x$ of functor
 $F$ at stage
${\ell A}$}, is called an element $x\in F({\ell A})$.

Now we element $r\in \r$ interpret as generalized element $i(r)\in_{\ell A}F$, where
$F=i(\r)$.
We have such elements so much how much loci. Transition to model
${\bf Set}^{\L^{op}}$  causes "reproduction" of element $r$. It begins
to exist in  infinite number of variants $\{i(r): i(r)\in_{\ell A}F,
\ell A\in \L \}$.

Note that since  4-metric $g^{(4)}$ is element of object
$\r^{\r^4\times \r^4}$ then "intuitionistic" 4-metric
begins
to exist in  infinite number of variants $i(g)^{(4)}\in_{\ell A}
i(\r^{\r^4\times \r^4})$.
Denote such variant as $i(g)^{(4)}(\ell A)$.

For simplification of interpretation we shall operate with objects of models
${\bf Set}^{\L^{op}}$. In other words, we shall write
$g^{(4)}(\ell A)$ instead of
$i(g)^{(4)}(\ell A)$.

Every variant $g^{(4)}(\ell A)$
of 4-metric $g^{(4)}$ satisfies to "own" Einstein equations \cite{4}
$$
R^{(4)}_{ik}(\ell A)-\frac{1}{2}g^{(4)}_{ik}(\ell A)[R^{(4)}(\ell
A)-2\Lambda(\ell A)]=
$$
$$
=\frac{8\pi G}{c^4}T_{ik}(\ell A).
$$
(Constants $c,G$ can also have different values at different stages $\ell A$).




It follows from  theory that when $\ell A=\ell C^\infty(\R^m)$ then
$$
g^{(4)}(\ell A)=[g\in_{\ell A}\r^{\r^4\times \r^4}]\equiv
g^{(4)}_{ik}(x^0,...,x^3, a)dx^idx^k,
$$
$$
a=(a^1,...,a^m)\in \R^m.
$$
Four-dimensional metric $g^{(4)}_{ik}(x^0,...,x^3, a)$ we extend to
(4+m)-metric
in space~$\R^{4+m}$
$$
g_{AB}^{(4+m)}dx^Adx^B\equiv
$$
\begin{equation}\label{met}
\equiv g^{(4)}_{ik}(x^0,...,x^3, a)dx^idx^k-da^{1^2}-...-da^{m^2},\\
\end{equation}
We get  $(4+m)$-dimensional pseudo-Riemannian geometry $<\R^{4+m},g_{AB}^{(4+m)}>$.

Symbolically procedure of creation of many-dimen\-sional variants of space-time geometry
by means of intuitionistic 4-geometry $<\r^4,g^{(4)}>$
 one can represent in the form of formal
sum
$$
g^{(4)}=c_0\cdot\underbrace{[g^{(4)}\in_{\bf 1}\r^{\r^4
\times \r^4}]}_{\mbox{\footnotesize 4-geometry}}+
$$
$$
+c_1\cdot\underbrace{[g^{(4)}\in_{\ell C^\infty(\R^1)}\r^{\r^4\times
\r^4}]}_{\mbox{\footnotesize 5-geometry}}+...
$$
$$
...+c_{n-4}\cdot\underbrace{[g^{(4)}\in_{\ell C^\infty(\R^{n-4})}
\r^{\r^4\times \r^4}]}_{\mbox{\footnotesize n-geometry}}+...,
$$
where coefficients $c_m$ are taked from the field of complex numbers.

Because number of stages is infinite,   we must
write integral instead of sum:
\begin{equation}\label{metr1}
  g^{(4)}=\int\limits_{\L}\D[\ell A]c(\ell A)[g^{(4)}
  \in_{\ell C^\infty(\R^{n-4})}\r^{\r^4\times
\r^4}].
\end{equation}
Use denotations of quantum mechanics \footnote{Dirac denotations:
 $|P\rangle=\psi(\xi)\rangle\equiv \psi(\xi)$; in given case $\psi(\xi)$ is
$g^{(4)}$ (representative
 of state $|P\rangle$), and
 $|P\rangle$ is $|g^{(4)}\rangle$ \cite[p.111-112]{5}.}:
$$
g^{(4)}\to |g^{(4)}\rangle, \
[g^{(4)}\in_{\ell C^\infty(\R^{n-4})}\r^{\r^4\times \r^4}]\to
|g^{(4)}(\ell A)\rangle.
$$
Then (\ref{metr1}) is rewrited in the form
\begin{equation}\label{metr2}
|g^{(4)}\rangle=\int\limits_{\L}\D[\ell A]c(\ell A)|g^{(4)}(\ell A)\rangle.
\end{equation}
Consequently, formal the Lawvere-Kock 4-geometry $<~\r^4,g^{(4)}>$ is
infinite sum
$$
\r^4=\int\limits_{\L}\D[\ell A]c(\ell A)\r^4_{\ell A}
$$
of classical many-dimensional pseudo-Riemmanian geometries
$\r^4_{\ell A}=$ $=<~\R^{4+m},g_{AB}^{(4+m)}(x,a)>$
every of  which
contains the foliation of 4-dimensional parallel universes (leaves) (under
fixing $a=const$).
Geometrical properties of these universes as it was shown in \cite{6,7}
to be different even within the framework of one stage~$\ell A$.

\smallskip
Now we  recall about environments of virtual reality
 which must appear under referencing to model of multiverse, in this instance,
 to model ${\bf Set}^{\L^{op}}$. This model is generator of virtual reality.
It is not difficult to understand that generalised element
$|g^{(4)}(\ell A)\rangle$ is  metric of concrete  environment
(=hyperspace $\r^4_{\ell A}$)
 with "number" $\ell A$.
 In other words, study of any object of theory $\T$ at stage $\ell A$
 is transition to one of the environments
 from repertoire of virtual reality generator ${\bf Set}^{\L^{op}}$.

\section{The G\"odel-Deutsch Multiverse}

As example of multiverse we consider
cosmological solution of Kurt G\"odel ~\cite{8}
\begin{equation}\label{mged}
g^{(4)}_{ik}=\alpha^2
\left(\begin{array}{cccc}
  1 & 0 & e^{x^1} & 0 \\
  0 & -1 & 0 & 0 \\
  e^{x^1} & 0 & e^{2x^1}/2 & 0 \\
  0 & 0 & 0 & -1
\end{array}\right).
\end{equation}
This metric satisfies the Einstein equations (\ref{ein}) with energy-momentum
tensor of dust matter
$$
T_{ik}=c^2\rho u_iu_k,
$$
if
\begin{equation}\label{gcond}
\frac{1}{\alpha^2}=\frac{8\pi G}{c^2}\rho, \ \ \Lambda=-\frac{1}{2\alpha^2}=
-\frac{4\pi G\rho}{c^2}.
\end{equation}
Take
\begin{equation}\label{const}
\alpha=\alpha_0+d,\ \Lambda=\Lambda_0+\lambda,\ \rho=\rho_0+\varrho,
\end{equation}
where $d,\lambda,\varrho\in D$ are infinitesimals and substitute these
in (\ref{gcond}). We get
$$
\frac{1}{(\alpha_0+d)^2}=\frac{1}{\alpha_0^2}-\frac{2d}{\alpha_0^3}=
\frac{8\pi G}{c^2}(\rho_0+\varrho),
$$
$$
2\Lambda_0+2\lambda=-\frac{1}{\alpha_0^2}+\frac{2d}{\alpha_0^3},\ \
\Lambda_0+\lambda=
-\frac{4\pi G\rho_0}{c^2}-\frac{4\pi G\varrho}{c^2}.
$$
Suppose that $\alpha_0,\Lambda_0,\rho_0\in \R$ are satisfied to
relations (\ref{gcond}). Then
$$
\lambda=-\frac{4\pi G}{c^2}\varrho, \ \
d=-\frac{4\pi G\alpha_0^3}{c^2}\varrho.
$$
Under interpretation in smooth topos ${\bf Set}^{\L^{op}}$ infinitesimal
 $\varrho\in D$ at stage
$\ell A=C^\infty(\R^m)/I$ is class of smooth functions of the
form $\varrho(a)\ mod\ I$,
where
$[\varrho(a)]^2\in I$ \cite[p.77]{3}.

Consider the properties  of the G\"odel-Deutsch multiverse at stage
 $\ell A=$ $\ell C^\infty(\R)/(a^4)$ \footnote{Here $(f_1,...,f_k)$ is
 ideal of ring
$C^\infty(\R^n)$ generated dy functions $f_1,...,f_k\in C^\infty(\R^n)$,
i.e. having the form
$\sum_{i=1}^kg_if_i$, where  $g_1,...,g_k\in C^\infty(\R^n)$ are arbitrary
smooth functions.}, where  $a~\in\R$. Obviously that it is possible to take
infinitesimal
 of form  $\varrho(a)=a^2$. Multiverse at this stage is 5-dimensional
hyperspace. This hyperspace contains a foliation, leaves of which are defined
by the equation $a=const$.
 The leaves are parallel universes in hyperspace (environment) $\r^4_
 {\ell A}$
with metric $g^{(4)}(\ell A)=g^{(4)}_{ik}(x,a)$ defined formulas
(\ref{mged}), (\ref{const}).
Density of dust matter $\rho=\rho_0+\varrho(a)$ grows from classical
value $\rho_0\sim 2\cdot 10^{-31}$ {\it g/cm$^3$} to $+\infty$
under $a\to \pm\infty$. Cosmological constant  grows  also infinitely
to $-\infty$.
Hence parallel universes  have different from our Universe
physical properties.

At stage $\ell A=\ell C^\infty(\R)/(a^2)$ \ $\varrho(a)=a$ and
$\rho=\rho_0+\varrho(a)\to
-\infty$
under $a\to -\infty$, i.e. $\rho$ is not phisically interpreted (we
have "exotic" matter
with negative density).

Finally, at stage ${\bf 1}=\ell C^\infty(\R)/(a)$ all
$\varrho(a)=d(a)=\lambda(a)=0$,
i.e. we have classical the G\"odel universe.

\section{The Friedman-Deutsch Multiverse}

Now we consider closed Friedman model of Universe, which in coordinates
$(x^0,\chi,\theta,\varphi),\ x^0=ct,$ has the following metric
$$
ds^2=g^{(4)}_{ik}dx^idx^k=
$$
\begin{equation}\label{fr1}
=c^2dt^2-R^2(t)[d\chi^2+ \sin^2\chi\ (d\theta^2+\sin^2\theta d\varphi^2)].
\end{equation}
This metric satisfies the Einstein equations with energy-momentum
tensor of dust matter
$$
T_{ik}=c^2\rho u_iu_k,
$$
under the condition that
\begin{equation}\label{fr2}
\rho R^3(t)=const=\frac{M}{2\pi^2}
\end{equation}

\begin{equation}\label{fr3}
\left\{
\begin{array}{l}
  R=R_0(1-\cos \eta),\\
  t=\frac{\displaystyle  R_0}{\displaystyle c}(\zeta-\sin\eta),
\end{array}\right.
\end{equation}
ãäå
\begin{equation}\label{fr4}
  R_0=\frac{2GM}{3\pi c^3},
\end{equation}
where
 $M$ is sum of body mass in 3-space \cite[p.438]{9}.

Let
\begin{equation}\label{fr5}
  G=k+d,\ \ \ d\in D
\end{equation}
where $k=6,67\cdot 10^{-8}\ [CGS]$ is classical gravitational constant.

At stage  ${\bf 1}=\ell C^\infty(\R)/(a)$ \ \  $d(a)=0$,
i.e. we have classical Friedman Universe.

Consider the  state of the Friedman-Deutsch multiverse at stage
 $\ell A=\ell C^\infty(\R)/(a^4),$
where  $a\in\R$. Obviously that it is possible to take
infinitesimal
 of form  $\varrho(a)=a^2$. Multiverse at this stage is 5-dimensional
hyperspace. This hyperspace contains a foliation, leaves of which are defined
by the equation $a=const$.
 The leaves are parallel universes in hyperspace (environment) $R^4_{\ell A}$
with metric $g^{(4)}(\ell A)=g^{(4)}_{ik}(x,a)$ defined formulas
(\ref{fr1})-(\ref{fr4}).

Radius of  "Universe" with number $a=const$ and dust density
as it follows from (\ref{fr2}) are equal to
$$
R=\frac{2}{3\pi
c^3}(k+a^2)(1-\cos \eta),
$$
$$
\rho(a)=\frac{27\pi c^3}{16k^3M^2(1-\cos \eta)^3}\left(1-\frac{3}{k}d(a)\right).
$$
So under $d=a^2$ the radius of parallel universes with numbers  $|a|\to+\infty$
grows to $+\infty$. The dust density $\rho(a)$
will decreas, then $\rho(a)$ is crossing zero and becomes negative, $\rho(a)\to -\infty$ under
 $|a|\to+\infty$.
All this says that parallel universes can have a physical characteristics which are
absolutely different from characteristics of our Universe.

\section{Transitions between parallel hyperspaces}

Change of stage $\ell A$ on stage $\ell B$ is morphism between two stages
$$
\ell B\stackrel{\Phi}{\rightarrow}\ell A.
$$
When $\ell A=\ell C^\infty(\R^n)$ and
$\ell B=\ell C^\infty(\R^m).$ Then transition $\Phi$ between stages
gives smooth mapping
$$
\phi:\R^m\ni b\to a\in \R^n,
$$
$$
a=\phi(b).
$$
Hence if constants $G=G(a), \Lambda=\Lambda(a)$ at stage $\ell A$,
then we have at new stage $\ell B$ \ \ $G=G(\phi(b)), \Lambda=\Lambda(\phi(b))$.
In other words, dependence of physical constants on exra-dimensions
is transformated in dependence of physical constants on some exra-field $\phi$.
This fact can be useful in connection with investigations,
concerning introduction effective fravitational constant depending on
some scalar field (see, for example \cite{10}).

\section{Conclusion}

As it follows from sections 4,5
the source of multiplicity of objects and appearance of  parallel hyperspaces
 are the physical constants (for example, $\rho, \Lambda, G$).
Reason this in following. Traditionally we  consider
physical constants as real numbers. It means
impossibility of findings of their exact values. So we musr take
that physical constant $K=K_0+d$, where $d$ is an infinitesimal.
The last gives multiplicity.


\end{document}